\documentclass{iaus}
\usepackage{graphicx}

\title[N/O-trends in Late-Type Galaxies] 
{N/O-trends in Late-Type Galaxies: AGB-stars, IMFs, Abundance Gradients and the Origin of Nitrogen}

\author[Mattsson]   
{Lars Mattsson}%
  
\affiliation{Department of Physics \& Astronomy, Uppsala University,
Sweden, \break email: mattsson@astro.uu.se}

\pubyear{2008}
\volume{254}  
\pagerange{1--6}
\date{?? and in revised form ??}
\jname{The Galaxy Disk in Cosmological Context}
\editors{J. Andersen, J. Bland-Hawthorn \& B. Nordstr\"om, eds.}
\begin{document}

\maketitle

\begin{abstract}
  Models of galactic chemical evolution (CEMs) show that the shape of the stellar initial mass function (IMF) 
  and other assumptions regarding star formation affect the resultant abundance gradients in models of late-type 
  galaxies. Furthermore, intermediate mass (IM) stars undeniably play an important role in the buildup of nitrogen 
  abundances in galaxies. Here I specifically discuss the nitrogen contribution from IM/AGB stars and how it affects 
  the N/O-gradient. For this purpose I have modelled the chemical evolution of a few nearby disc galaxies using different IMFs and star 
  formation prescriptions. It is demonstrated that N/O-gradients may be used to constrain the nitrogen contribution 
  from IM/AGB-stars. 
\keywords{Galaxies: abundances, evolution, spiral, Stars: AGB and post-AGB}
\end{abstract}

\firstsection 
\section{Introduction}
Much progress has been made in modelling nucleosynthesis in stars in general, but the origin of nitrogen still remains somewhat uncertain. In the 
classical picture, the apparent increase in nitrogen production with metallicity, as inferred from spectroscopy of galactic and extragalactic 
HII-regions, is a manifestation of nitrogen being either primary or secondary (produced from primordial elements or with 
heavier elements as "seeds"). However, reality often turns out to be more complex than our simplest models of it. 
The origin of nitrogen is probably no exception.

During the last few years, reasonably accurate stellar nitrogen abundances have become available also for Population II stars 
(see, e.g., \cite{Cayrel04, Ecuvillon04, Israelian04, Spite05}). The rather large nitrogen abundances relative to iron seen in halo stars in the 
Milky Way suggest that a relatively large fraction of the nitrogen should come from Type II supernovae, especially at low metallicity. However, recent 
stellar evolution models, including nucleosynthesis, all appear to predict significant nitrogen production in intermediate-mass (IM) stars 
(\cite{Marigo01, Izzard04, Gavilan05, Karakas07}). The typical production factor would be of the order $\sim 10^{-3}$ of the initial 
stellar mass. If IM-stars of all metallicities produce that much nitrogen, it would require that metal-poor galaxies (or regions within a galaxy) with 
low N/O-ratios have very young stellar populations, in fact, they have to be so young that IM-stars have not yet made any significant contribution to the 
interstellar nitrogen abundance of these objects, unless the production of oxygen is much greater at low metallicity. 

In this work I have considered radial gradients of O/H and N/O in the Milky Way and five other nearby late-type galaxies in order to find additional
clues to the origin of nitrogen. The advantages of considering the abundance gradients within galaxies, rather than over-all trends obtained from 
the global properties of galaxies, are that the same stellar nucleosynthesis prescriptions (yields) must be able to simultaneously reproduce the 
radial abundance gradients in galaxies with quite different properties (sizes, total masses, gas distributions, etc.) and the very same yields must
also be consistent with the fact that galaxies with similar O/H-gradients may have quite different N/O-gradients. Models of chemical evolution (CEMs)
may thus be able to provide insight and constraints on stellar yields when several galaxies are modelled and analysed in parallel.

\section{Data}
The oxygen and nitrogen abundances are taken from \cite[Pilyugin et al. (2003, 2004)]{Pilyugin01a, Pilyugin01b}. which provide a large compilation of 
data that is consistent in the sense that all abundances are derived using the same strong-line calibration (P-method, see 
\cite{Pilyugin01a, Pilyugin01b}). Data for HI and H$_2$ (CO) were collected from the literature (see Table \ref{params} for references). To obtain a 
consistent and, possibly, a more correct set of data for the gas distributions, the conversion from CO-flux to H$_2$ surface density has been redone 
using the oxygen-dependent calibration by \cite[Wilson (1995)]{Wilson95}.

  \begin{table}
  \begin{center}
  \caption{\label{params} Model parameters and adopted distances. See Sect. \ref{models} for further details.}
  \begin{tabular}{lllllllll}
  \hline
  Galaxy     & $D$ & $R_{\rm d}$ & $M_{\rm d}$ & $\eta$ & $\tau_0$ & $\tau_1$ & $m_{\rm up}$ & Refs. for CO and HI data:\\
            & Mpc & kpc         & $10^{10} M_\odot$ &    & Gyr      & Gyr             & $M_\odot$ & \\
  \hline
  Milky Way	& 0.00 & 2.8 & 6.50 & 1.0 & -1.3     & 2.9 & 60/70 & Sanders et al. (1984)\\
  NGC 224	& 0.69 & 6.5 & 12.5 & 3.0 & -5.0     & 8.5 & 60/70 & Koper et al. (1991)\\				
  NGC 598	& 0.79 & 2.5 & 1.50 & 1.0 & $\infty$ & -   & 60/70 & Corbelli (2003)\\		
  NGC 2403	& 3.30 & 2.8 & 2.00 & 1.0 & $\infty$ & -   & 60/70 & Wevers et al. (1986), \\
            &       &    &      &     &          &     &       & Thornley \& Wilson (1995)\\			
  NGC 3031	& 3.25 & 2.7 & 8.00 & 2.0 & -5.5     & 5.0 & 60/70 & Rots et al. (1975)\\				
  NGC 6946	& 5.50 & 4.0 & 10.0 & 1.0 & $\infty$ & -   & 60/70 & Tacconi \& Young (1986)\\
  \hline       
  \end{tabular}
  \end{center}
  \end{table}

\section{Models}
\label{models}
I present detailed, numerical CEMs for the Milky way and five other nearby spirals, computed using the code described in detail in 
\cite[Mattsson (2008)]{Mattsson08}. Since the halo contributes very little to the present-day abundance gradients, except for at very large galactic 
radii, where the disc density becomes comparable with the halo density (see, e.g., \cite{Chiappini01}), the model is constructed as a single exponential 
disc, formed by exponentially decaying baryonic infall, i.e., 
\begin{equation} 
\dot{\Sigma}_{\rm inf.}(r,t) = {\Sigma_{\rm d}(r,t_0)\over\tau(r)} \left\{\exp\left[-{t_0\over \tau(r)}\right]\right\}^{-1}\exp\left[-{t\over\tau(r)}\right], 
\end{equation}
where $\tau$ is the infall-time scale and $t_0$ denotes present time. The star formation rate is prescribed by a modified "Silk-law" 
(\cite{Wyse89, Bossier03}), i.e., 
\begin{equation}
\dot{\Sigma}_\star(r,t) = \eta\, \dot{\Sigma}_{\star, {\rm MW}}(R_\odot,t_0)\,{\Omega_0(r)\over \Omega_{0,{\rm MW}}(R_\odot)}\,\left[{\Sigma_{\rm gas}(r,t)\over \Sigma_{\rm gas}(R_\odot,t_0)} \right]^{1+\varepsilon},
\end{equation}
where $\Omega_0$ is the present-day ($t=t_0$) angular frequency of the disc (taken from \cite{Sofue99}) and $\eta$ is a 
parameter regulating the star-formation efficiency relative to the Milky Way. To simulate the inside-out formation of a 
galactic disc, the infall time scale is assumed to have a linear radial dependence, $\tau(r) = {\rm max}(0, \tau_0 + \tau_1\, r/r_{\rm d})$,
where $\tau_0, \tau_1$ are arbitrary constants and $r_{\rm d}$ is the disc scale length. For $\tau_0 = -1.3$, $\tau_1 = 2.9$ and $r_{\rm d} = 2.8$ 
kpc, the above relation is similar to that used by \cite[Romano et al. (2000)]{Romano00} and \cite[Boissier \& Prantzos (1999)]{Bossier99}. For the 
gas-rich galaxies (NGC 598, NGC 2403, NGC 6946), the infall rate is assumed to be constant 
at all radii, indicated by $\tau\to\infty$ in Table \ref{params}.

The nucleosynthesis prescriptions were taken from \cite[Chieffi \& Limongi (2004)]{Chieffi04} for high mass stars and 
\cite[van den Hoek \& Groenewegen (1997)]{vandenHoek97} and \cite[Gavil\'an et al. (2005)]{Gavilan05} for low and intermediate stars (see Fig. 
\ref{yields}). The latter set of yields has now been extended to low metallicities (\cite{Gavilan07}). 

  \begin{figure}
  \begin{center}
  {
  \includegraphics[width=44mm]{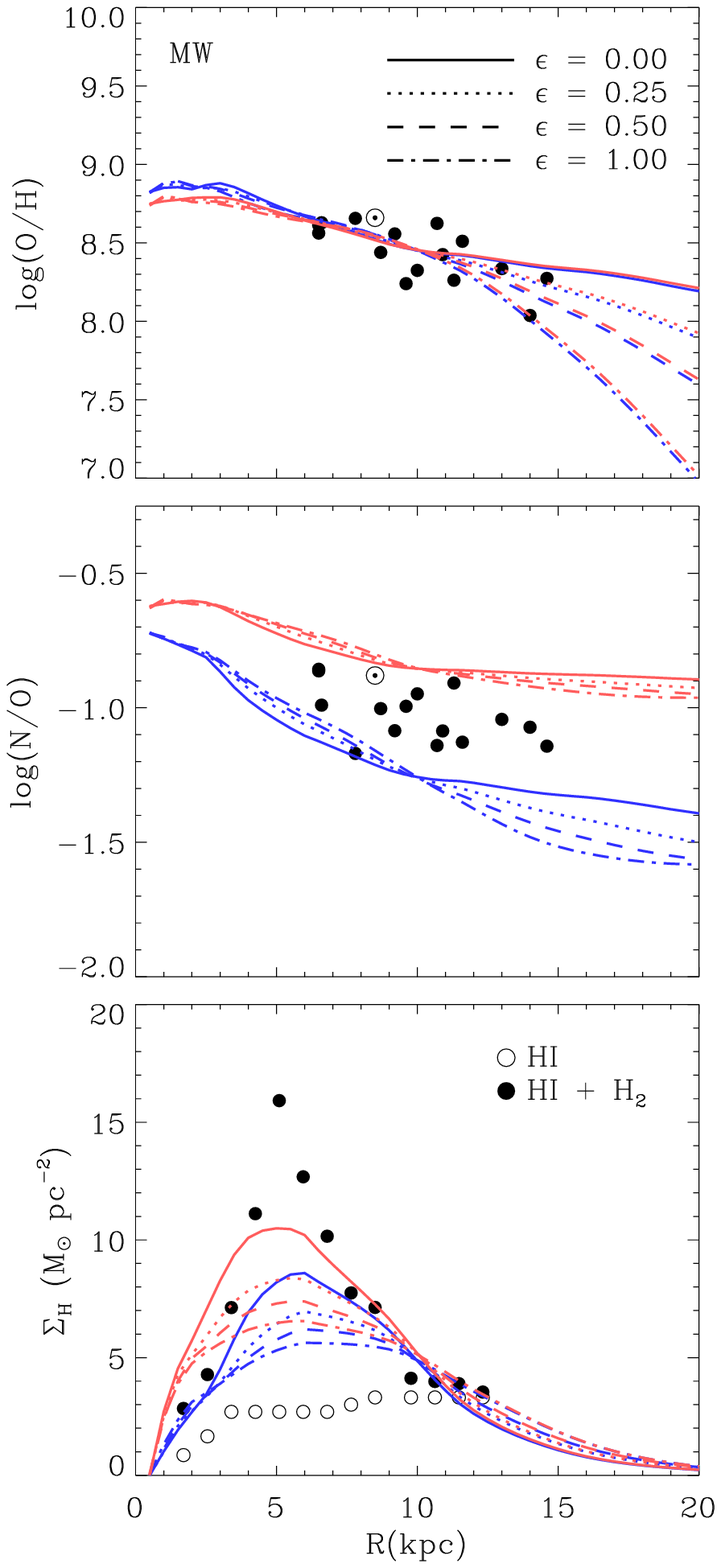}
  \includegraphics[width=44mm]{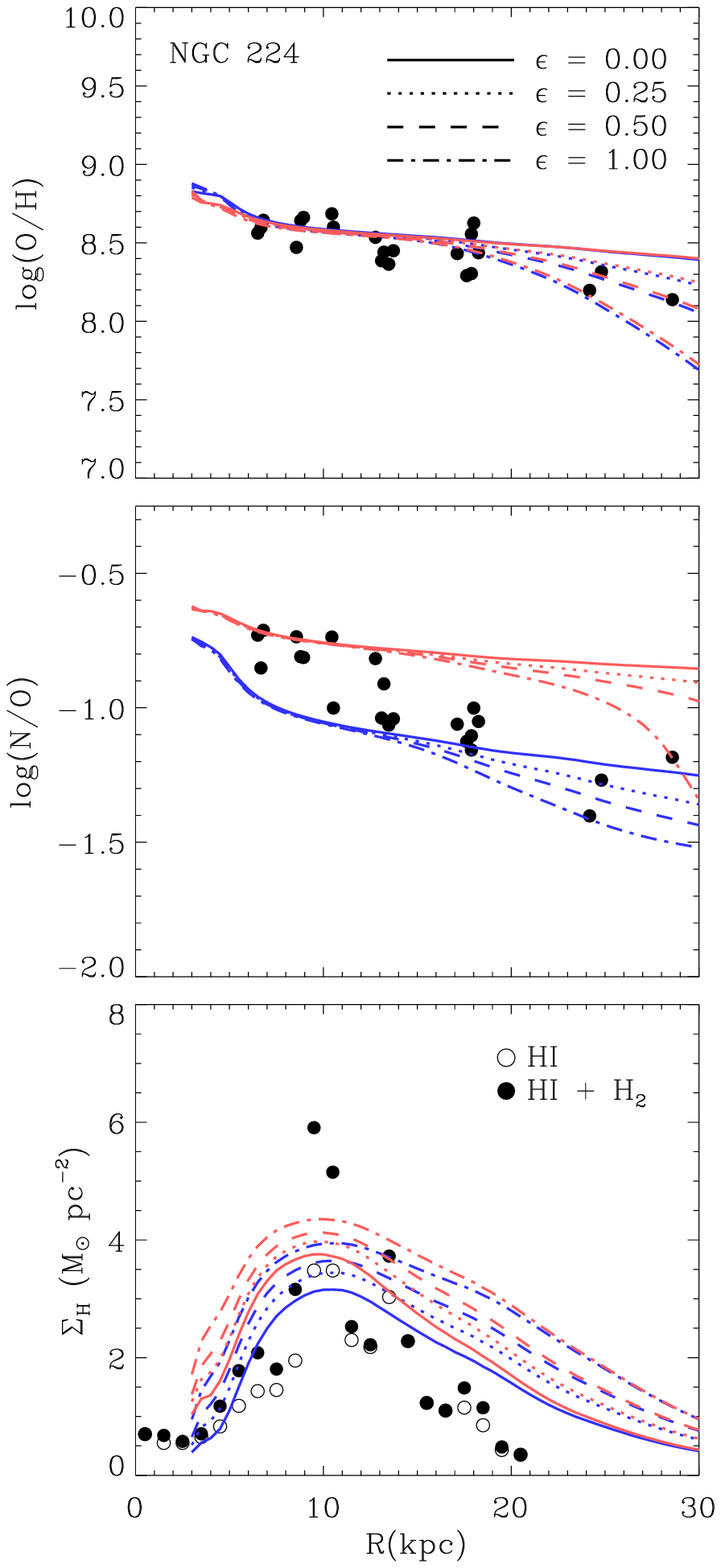}
  \includegraphics[width=44mm]{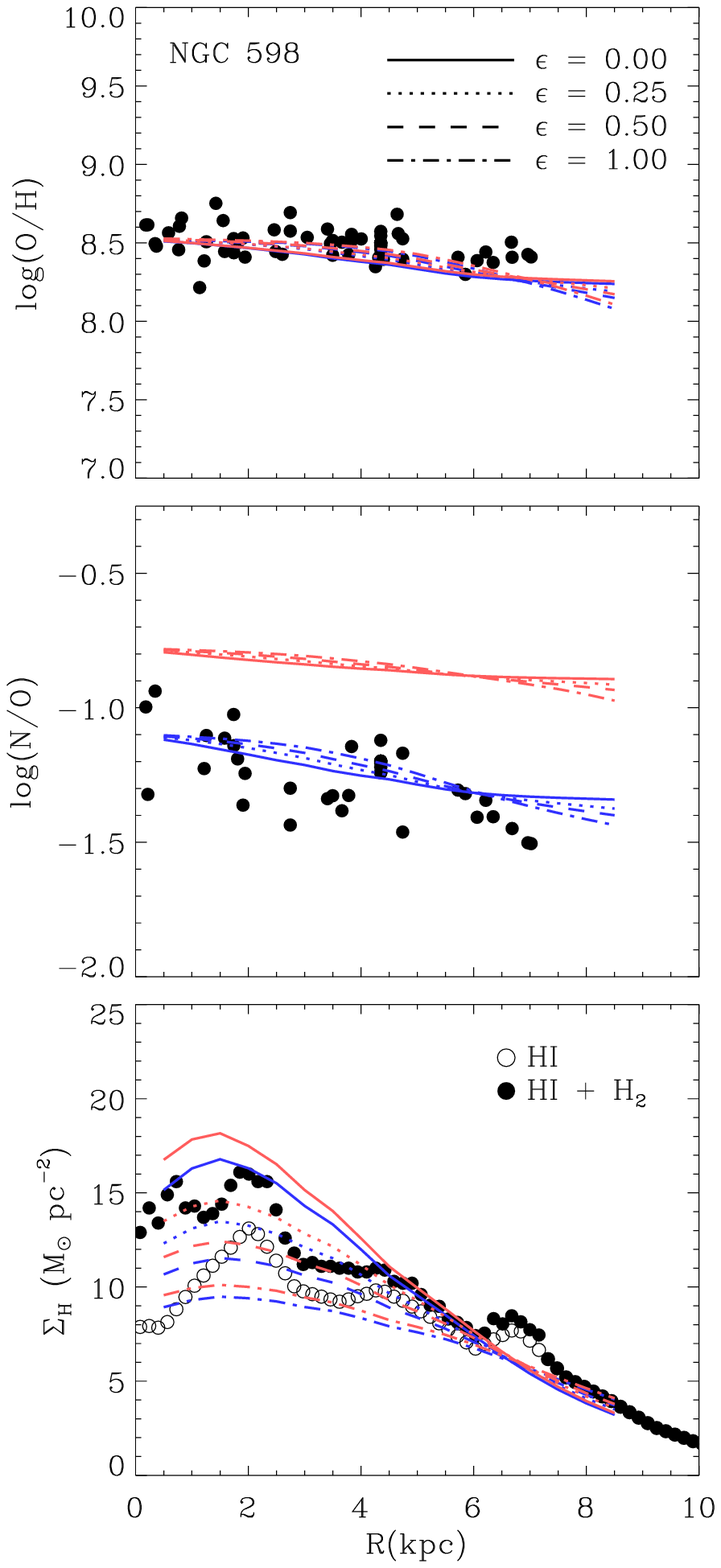}
  }
  \end{center}
  \caption{\label{result1}
  Radial trends for O/H (upper panels), N/O (lower panels) and the surface density of hydrogen. Black dots show the
  abundances from Pilyugin et al. (2003, 2004), red lines show models with yields from Gavil\'an et al. (2005, 2007) and blue lines show
  models with yields from van den Hoek \& Groenewegen (1997). The solar symbols show the solar values according to Asplund et al. (2005).}
  \end{figure}
 
\section{Results and Discussion}
\begin{figure}
  \begin{center}
  {
  \includegraphics[width=44mm]{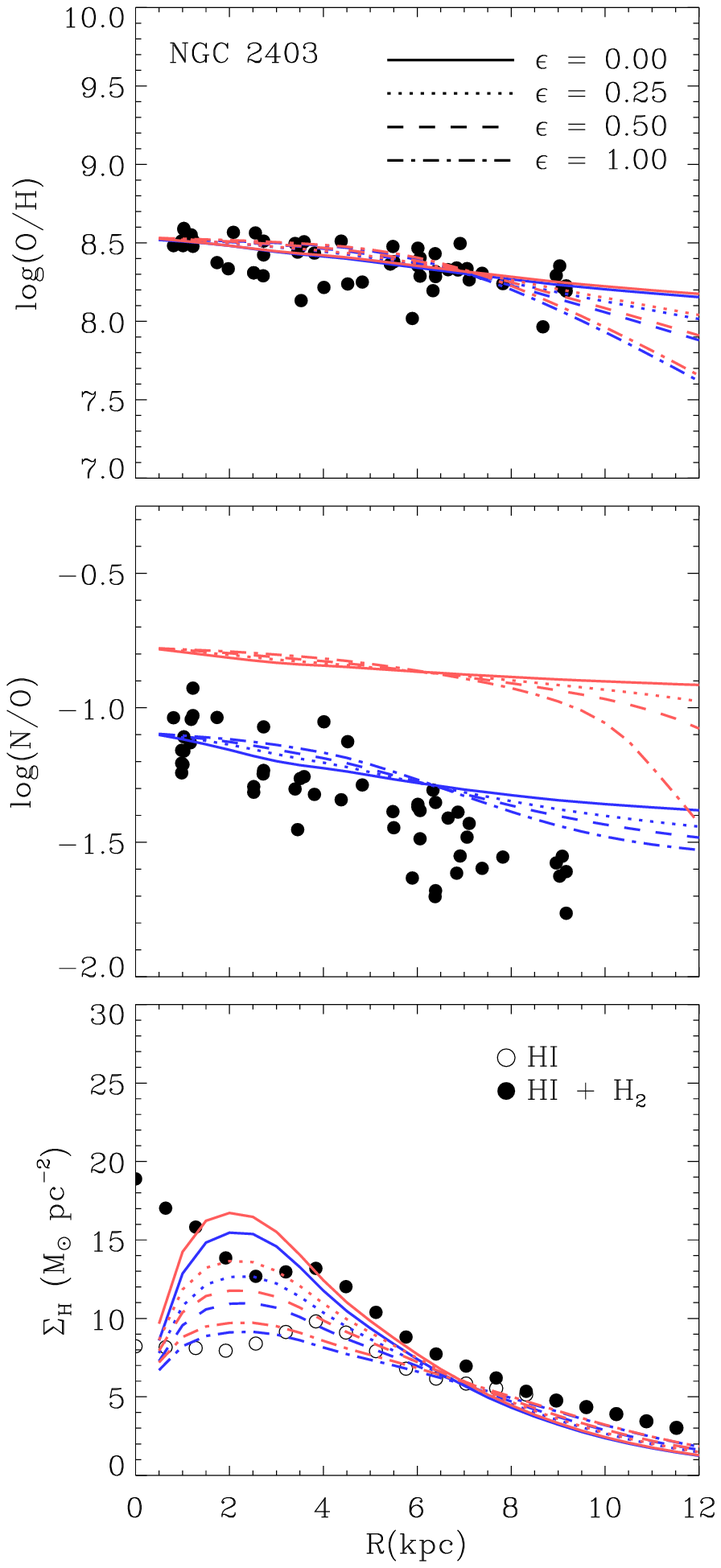}
  \includegraphics[width=44mm]{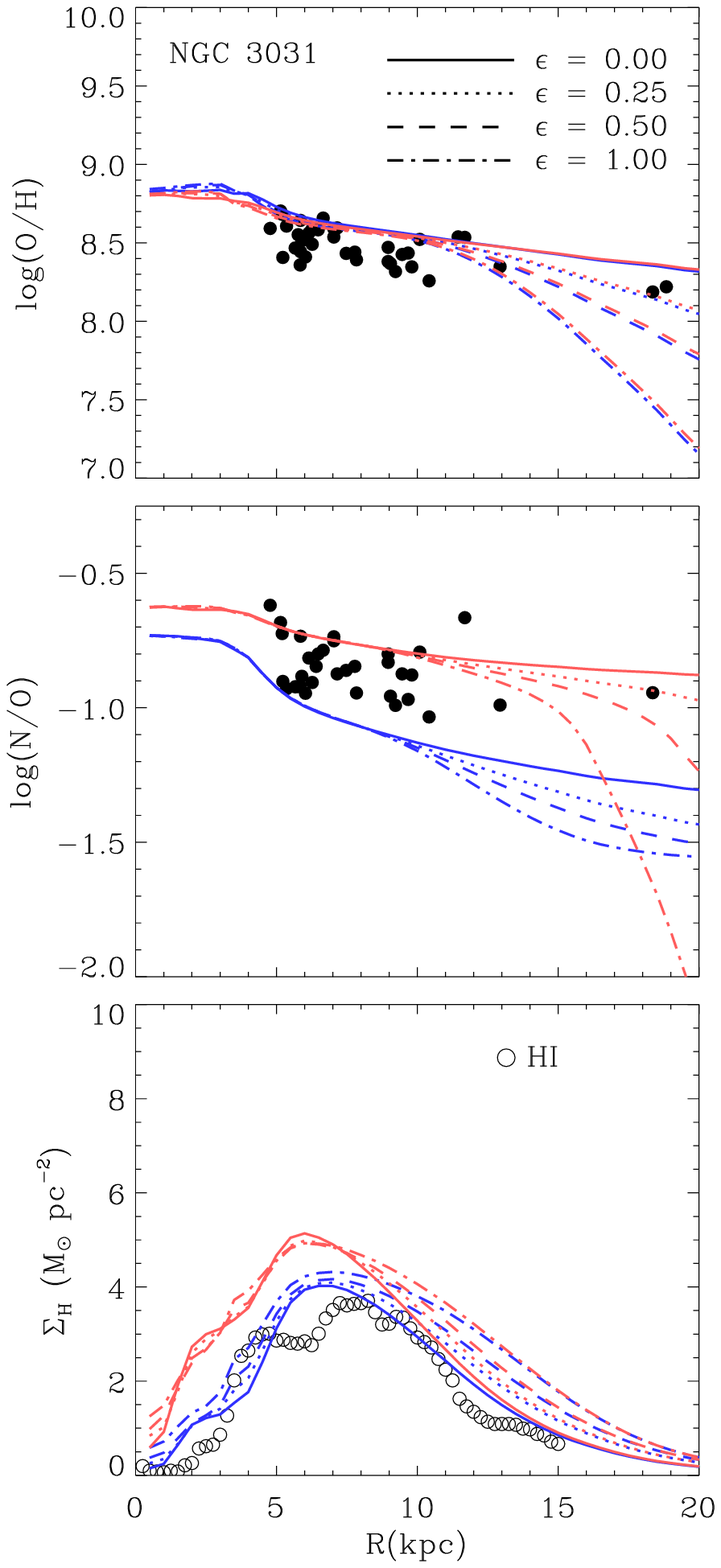}
  \includegraphics[width=44mm]{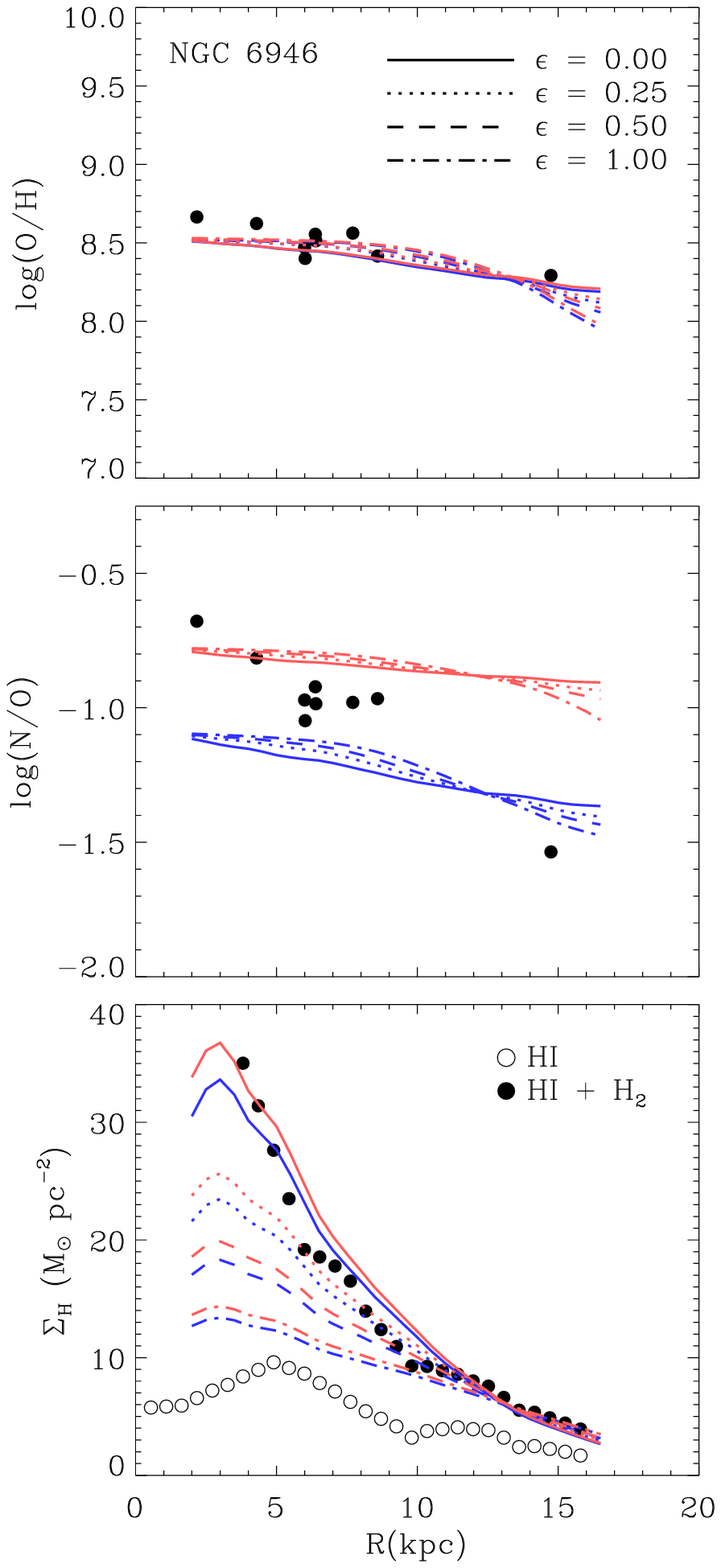}
  }
  \end{center}
  \caption{\label{result2} 
  Same as Fig. \ref{result1}, but for NGC 2403, NGC 3031 and NGC 6946.}
  \end{figure}

\begin{figure}
  \begin{center}
  {
  \includegraphics[width=66mm]{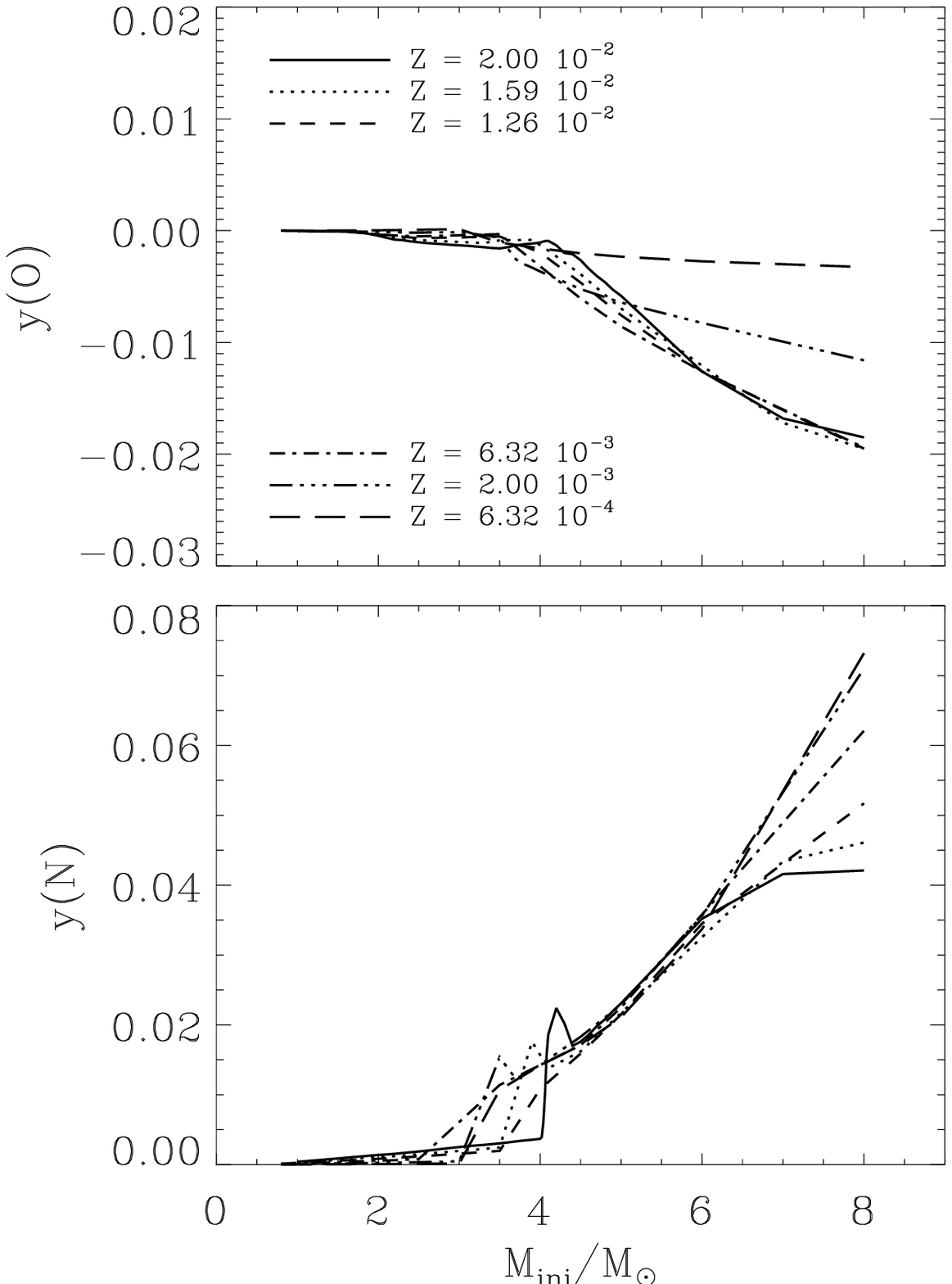}
  \includegraphics[width=66mm]{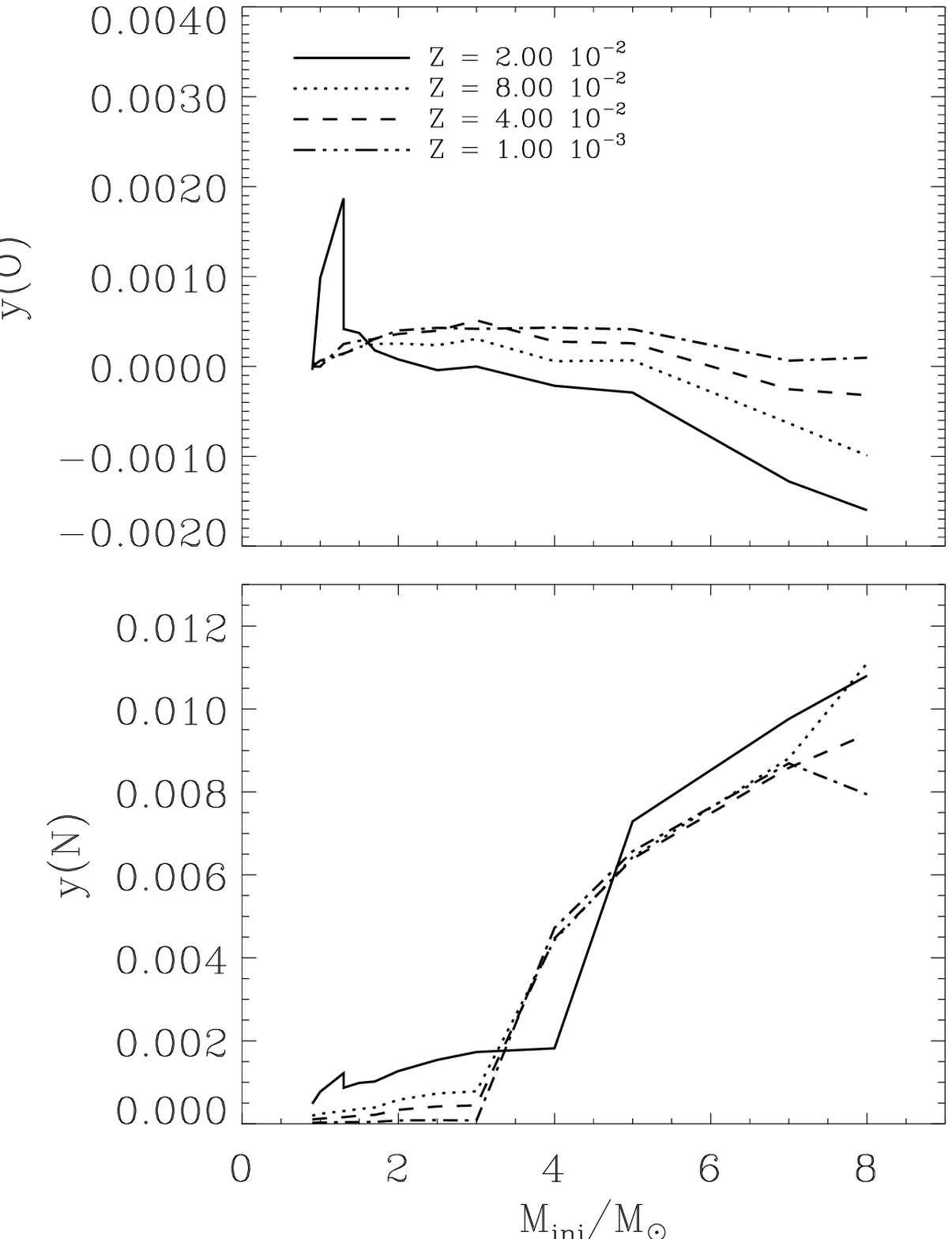}
  }
  \end{center}
  \caption{\label{yields} 
   Left panels: N- and O-yields from Gavil\'an et al. (2005, 2007) including the previously unpublished sets for low metallicity (left).
   Right panels: the same from van den Hoek \& Groenewegen (1997). Note that the N-yields by Gavil\'an et al. are roughly five times larger 
   than those by van den Hoek \& Groenewegen.}
  \end{figure}

The modelling result using a \cite[Scalo (1986)]\cite{Scalo86} IMF is presented below in terms of O/H- and N/O-gradients and radial hydrogen distributions (Fig. 
\ref{result1} and Fig. \ref{result2}). To avoid over-production of oxygen the IMF was truncated at $m_{\rm up}=60 M_\odot$ for models with yields 
from van den \cite[Hoek \& Groenewegen (1997)]{vandenhoek97} and at $m_{\rm up}=70 M_\odot$ for models with yields from 
\cite[Gavil\'an et al. (2005, 2007)]{Gavilan05, Gavilan07}. 

I adjusted the remaining model parameters so that the O/H-gradients were reproduced and a reasonable agreement with the observed hydrogen
distributions were obtained. For the Milky Way, the properties of the solar neighbourhood were also used as constraints. The modelled N/O-gradients 
are then not in very good agreement with the observations - the only exceptions 
being NGC 598 and NGC 3031, where the models reproduce the N/O-gradient reasonably well, although with different sets of yields. 
If the abundances can be trusted, which is very likely, this indicates a relatively strong correlation with metallicity of the integrated nitrogen yields. 
In contemporary stellar yields for IM-stars (see Fig 1.) there is no strong dependence on metallicity, which perhaps indicates that the IMF is varying. 
Note, also, that newer nitrogen yields, in general (see \cite{Gavilan05}), are larger than those by 
\cite[van den Hoek \& Groenewegen (1997)]{vandenHoek97}.

\begin{figure}
  \begin{center}
  \includegraphics[width=132mm]{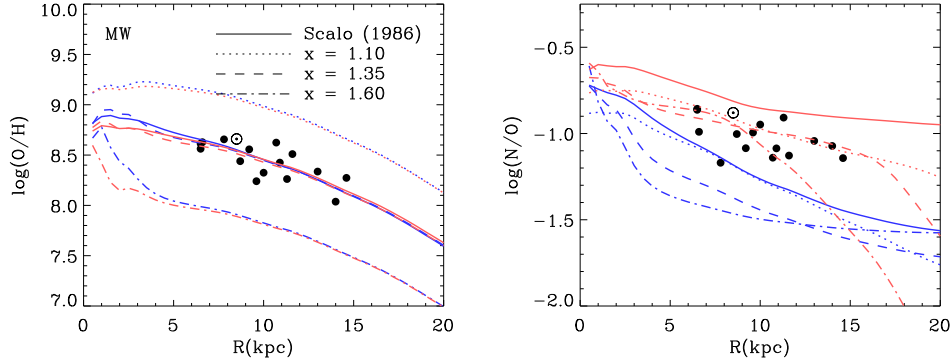}
  \end{center}
  \caption{\label{imf}
  Variations of the IMF. Changing the slope of a simple power-law IMF ($m_{\rm up} = 50 M_\odot, $ $m_{\rm low}= 0.08 M_\odot$ 
  are left unchanged) will shift the O/H-gradient up or down (see left panel), but leave $d({\rm O/H})/d{\rm R}$ almost unaffected.
  A Salpeter (1955) IMF ($x = 1.35$) which provides an acceptable fit to the observed N/O-ratios (right panel).}
  \end{figure}

Variations of the slope of the IMF changes the N/O-gradient, but at the same time the over-all oxygen abundance will change (see Fig. \ref{imf}).
A \cite[Salpeter (1955)]{Salpeter55} IMF, $\phi(m)\sim m^{-(1+x)}, x = 1.35$, truncated at $m_{\rm low}=0.08 M_\odot$ and $m_{\rm up}=50 M_\odot$ gives 
a result similar to that with the \cite[Scalo (1986)]{Scalo86} IMF for the O/H-gradient, but also a reasonable fit to the observed N/O-gradient for the 
Milky Way (see Fig. \ref{imf}) using the \cite[Gavil\'an et al. (2005, 2007)]{Gavilan05, Gavilan07} yields. However, for the other five galaxies, this 
combination of IMF and yields does not really improve the fit to the observed abundances. Hence, resolving the O/H-N/O-inconsistency by changing the IMF 
requires that the IMF varies over the galactic disc, or, effectively, from galaxy to galaxy.

A stellar-metallicity dependence in the nitrogen yields of IM stars is also a reasonable explanation. Since the N/O-gradients predicted by the 
models are too flat in three cases (NGC 224, NGC 2403 and NGC 6946), the yields may need to increase with metallicity. If the N-yields are as 
high as those by \cite[Gavil\'an et al. (2005)]{Gavilan05} for IM-stars around solar metallicities, but as low as those by 
\cite[van den Hoek \& Groenewegen (1997)]{vandenHoek97} at subsolar metallicities, the observed slopes might be reproduced. On the other hand, 
it is quite likely that such an {\it ad-hoc} modification of the yields may lead to gradients that are too steep for the Milky Way and NGC 3031. 

\section{Conclusions}
Employing a standard IMF (\cite{Scalo86}) with a high-mass cut-off at $m_{\rm up}=60/70 M_\odot$ the O/H-gradients are well-reproduced by numerical
CEMs. However, the N/O-gradients cannot, in general, be reproduced simultaneously. This inconsistency can be explained in several ways, e.g., 
\begin{enumerate}
\item the production of nitrogen in intermediate-mass stars is strongly correlated with metallicity,
\item the IMF varies over the galactic disc (or, effectively, from galaxy to galaxy) such that the N/O-ratio is affected, 
\item the infalling gas that forms the disc is not primordial, but significantly enriched with heavier elements (from Pop. III or Pop. II stars?), 
\item or the P-method produces systematic errors.
\end{enumerate}
I believe that the explanations (a) and (b) are more likely then the other two. The main reasons are: (1) many ingredients of stellar 
evolution models that affects the nucleosynthetic yields (e.g., mass loss rates and the efficiency of hot-bottom burning) are still
highly uncertain, and (2) the turn-over in the IMF at low stellar masses appears correlated with the Jeans mass (\cite{vanDokkum08}), 
(3) significant effects from chemically enriched infall is only expected in the outer parts of the discs, where the abundances are low,
and (4) the P-method is in very good agreement with the standard electron-temperature method (\cite{Pilyugin01a, Pilyugin01b}).
Furthermore, the stellar yields and detailed properties of the IMF are probably the greatest uncertainties in CEMs in general. 

\begin{acknowledgments}
Leonid Pilyugin is thanked for supplying the abundance data and explaining some issues regarding
the P-method. Marta Gavil\'an and collaborators are thanked for making their low-metallicity yields available.
\end{acknowledgments}

\end{document}